# Using Unguided Peer Collaboration to Facilitate Early Educators' Pedagogical Development: An Example from Physics TA Training

Apekshya Ghimire * and Chandralekha Singh

Department of Physics & Astronomy, University of Pittsburgh, Pittsburgh, PA and 15260, USA; clsingh@pitt.edu
* Correspondence: apg61@pitt.edu

**Abstract**

Many early career educators, such as teaching assistants (TAs) in college courses, as well as pre-college educators, need help both with content and pedagogical knowledge to effectively help their students learn. One pedagogical approach that has been found effective in prior studies is collaboration with peers. Collaborative learning not only has the potential to help educators develop content knowledge but can also improve their pedagogical knowledge. This study examines the performance of physics graduate students, enrolled in a professional development course for teaching assistants (TAs), on the Magnetism Conceptual Survey, highlighting the impact of peer collaboration on learning both content and pedagogy. Peer interaction significantly improved performance, driven by both construction of knowledge (where the group answered a question correctly but only one member had the correct individual response) and co-construction of knowledge (where the group succeeded despite both members initially answering incorrectly). Beyond improving content understanding, peer collaboration can also foster pedagogical skills by encouraging early educators such as TAs to use peers as learning resources and communicate ideas effectively to support mutual understanding. These dual benefits—enhancing both content mastery and teaching abilities—demonstrate that this approach holds value not only for the professional development of TAs but can also be adapted for pre-college professional development programs to improve teaching and learning outcomes.

## 1. Introduction and Framework

*1.1. Value of Professional Development for Educators*

Many early career educators, e.g., teaching assistants (TAs) in college courses and pre-college educators, often enter instructional roles with limited formal training in pedagogy. These instructors require professional development (PD) that supports both content knowledge and pedagogical skills to effectively facilitate student learning. However, limited instructor time and institutional resources often constrain the depth and scope of formal PD programs. In the context discussed in this research, professional development refers to structured learning experiences that enhance TAs' knowledge of physics and their ability to facilitate student learning through research-informed teaching practices.

*1.2. Pedagogical Development Through Peer Collaboration*

Research suggests that peer collaboration is valuable for professional development of educators, and it is an effective pedagogical approach that educators can employ in their own courses [1-17]. However, for educators to recognize its value and implement collaborative learning methods in their own classrooms, they must first experience its value for their own learning [18-28]. Incorporating collaborative learning into teacher professional development programs can help educators enhance their pedagogical knowledge and teaching practices while also deepening their understanding of the subject matter [29-31]. The dual benefits of improving both content mastery and teaching skills underscore the value of this approach for professional development, not only for graduate teaching assistants discussed here but also for pre-college educators, to enhance their teaching and learning outcomes.

In this article, we discuss the benefits of making physics teaching assistants (TAs) in a professional development course collaborate with peers to improve both their content and pedagogical knowledge. The development of content and pedagogical knowledge described here does not require professional development leaders to invest their time and provide input while TAs collaborate. Considering professional development leaders have limited time to provide feedback to early educators, the type of professional development described in this article can be adapted for peer educators' communities of practice (CoPs, in which like-minded educators support each other) [32]. A community of practice consists of people who "share a concern or a passion for something they do and learn how to do it better as they interact regularly" [33]. The CoP concept was initiated by Lave, then expanded upon by Wenger, and can emerge from members' shared interests and goals [34]. By participating in a CoP, members exchange knowledge, learn from one another, and shape their individual identities [34]. In a CoP [34], educators can be encouraged to connect with peers either in person within their local communities or virtually through platforms like Zoom to engage in collaborative activities that strengthen their understanding and application of collaborative learning strategies [35].

*1.3. Unguided Peer Collaboration*

In this study, we focus on unguided peer collaboration, defined as peer-to-peer interaction in which TAs work together to solve content-related problems without guidance or input from an instructor or facilitator. It allows TAs to practice co-constructing knowledge, articulating ideas, and enhance their confidence in using collaborative methods with their students. Research shows that students retain knowledge better when they engage in group work rather than working alone [36]. These findings align with the frameworks of distributed cognition and the zone of proximal facilitation model [36, 37], which suggest that collaborative work enhances learning. Students can benefit from working together when peers help recognize relevant prior knowledge and address uncertainties. In situations where students struggle to tackle a problem independently, combining their knowledge allows them to arrive at the correct solution. Therefore, promoting out-of-class collaboration can be particularly valuable when classroom time is limited, as it helps students build a deeper understanding of core concepts. This approach also enables instructors to concentrate on more challenging ideas that fall outside the students' current zone of proximal development (ZPD), providing targeted scaffolding in those areas [38].

Peer collaboration also fosters growth in other important areas, such as skills in scientific communication and collaboration. Students are frequently more at ease sharing their ideas and voicing doubts with peers rather than instructors, as they may feel more at ease questioning their peers' reasoning rather than confronting an authority figure. This encourages confidence and offers chances to practice scientific communication and critical thinking [39-45]. Working with peers in an equitable learning environment can reduce anxiety, allowing students to explain their reasoning without fear of judgment, freeing up cognitive resources that would otherwise be spent on managing anxiety [46-54]. Additionally, peer collaboration can positively influence motivational beliefs, such as self-efficacy [43, 55, 56], which has been shown to correlate with improved performance in STEM courses [57-59].

In addition to using context-rich problems for collaborative group problem-solving [60, 61], Mazur's peer instruction approach has proven to be highly effective in college-level physics courses [62-64]. This approach combines lectures with think-pair-share tasks, where students work on multiple-choice questions related to physics concepts [62, 65]. As a formative assessment method, it fosters student learning by making them accountable during peer discussions, where they are required to explain their reasoning on various physics topics [62, 63, 65]. Research indicates that self-efficacy is crucial in this process, as students strengthen their understanding by explaining their ideas to each other [63]. While studies involving groups of three or more students [60, 61, 66, 67] suggest assigning roles like group leader or timekeeper can be beneficial, other research emphasizes the value of allowing students to select their partners when collaborating in pairs, as familiarity with peers can enhance the learning experience [68, 69].

*1.4. Relevance to Teaching and Study Motivation*

Like many other disciplines [70-75], in physics courses, students must simultaneously develop a deep conceptual understanding of the subject matter and effective problem-solving skills [76-78]. Peer collaboration, within and beyond the classroom, is an important tool to help achieve these goals. Research studies indicate that providing students with opportunities to learn from their peers, alongside support from course instructors, can be an important strategy to enhance their understanding [60, 61, 79]. Dewey's framework for participatory democracy emphasizes providing students with a supportive environment where they can collaborate and develop intellectually [80], while Hutchins highlights the value of distributed cognition [37], where collaborative group work helps expand cognitive resources and optimize outcomes. The Zone of Proximal Facilitation (ZPF) model [36], which builds on Vygotsky's Zone of Proximal Development (ZPD), further supports this by predicting that students with some knowledge of the content but unable to complete tasks individually can succeed through collaboration and learn, leveraging their collective expertise [81].

Previous research from our group underscores the significant benefits of peer collaboration in physics education. Studies using physics surveys demonstrate that group performance consistently surpasses individual performance across various levels, including among introductory physics students [82, 83] and graduate students [84]. This collaborative advantage extends to advanced topics like quantum mechanics [85]. Moreover, peer collaboration promotes long-term knowledge retention, as students often retain their understanding when reassessed individually after group discussions [86] Compared to students working alone, those who collaborated showed marked improvement in their understanding. Furthermore, students co-constructed knowledge during peer interactions, even without instructor guidance with approximate rates ranging from 20% to 30% [83-85]. Co-construction of knowledge occurs when students, who initially do not have the correct solution individually, work together to arrive at the right answer. This can happen when students with the same or different incorrect answers discuss their reasoning and identify flaws in their approaches. Even if both students share the same incorrect answer, they may recognize uncertainties in their reasoning and collaborate to refine their approach, ultimately reaching the correct solution through co-construction of knowledge.

Inspired by these frameworks and findings to provide the benefits of learning content as well as pedagogy, our study examines the impact of unguided peer collaboration on TAs' performance on the Magnetism Conceptual Survey (MCS) [87] in a TA professional development course for these early educators in introductory physics courses. The MCS is designed for introductory physics students in college or high school, i.e., its content is highly relevant to teaching at the high school and early college levels. In this study, the MCS served as a context to provide TAs with the opportunity to engage in collaborative learning around content they are likely to teach, enabling us to explore how they interact, reason and support one another during unguided peer discussions.

Thus, we analyze their individual and group performances to assess how peer collaboration affects their understanding of magnetism concepts, especially in the context of preparing them for teaching in which they can harness the benefits of collaborative learning pedagogy. Our research, which includes groups of two TAs (and occasionally three), allows participants to choose their partners after individually completing the MCS. By comparing their individual and collaborative performances, we determine the extent to which peer collaboration enhances TAs' conceptual knowledge of magnetism while improving their pedagogical readiness for teaching introductory physics since TAs learn the pedagogy of collaborative learning.

Our findings have direct implications for teacher preparation and professional development programs for early educators. Incorporating structured peer collaboration activities in such programs, even in unguided formats, can be an effective strategy for improving their teaching outcomes both due to improved content and pedagogical knowledge. Considering professional development leaders have limited time, apart from the early college educators such as TAs discussed here, teachers-in-training or early career educators of pre-college courses can also be encouraged and supported to use peer collaboration to deepen their content understanding and refine their instructional techniques. Facilitating discussions in a supportive CoP (Community of Practice) where participants actively engage in construction and co-construction of knowledge can help STEM educators not only learn challenging concepts but also develop understanding of pedagogy and how to foster collaborative learning environments for their students. Thus, integrating peer collaboration into professional development initiatives has the potential to enhance teaching and learning outcomes across educational contexts.

## 2. Research Questions

While the primary instrument used in this study, the Magnetism Conceptual Survey (MCS), was designed to assess physics content knowledge, we do not use it as a measure of TA conceptual mastery. Instead, the MCS serves as a contextual tool to stimulate meaningful peer interaction, reasoning, and collaborative engagement with material TAs are likely to teach. Our focus is not solely on content gains but on how TAs collaboratively engage in construction and co-construction of knowledge, practices that are critical to pedagogical development. The learning processes while collaborating and TAs' reflections on their collaborative learning experiences provide insights into their evolving pedagogical thinking and their confidence in implementing collaborative methods in future instruction.

While TAs learning the pedagogy of collaborative learning is an implicit goal of the study, the study explored the following research questions to investigate the effect of working with others on TAs' performance as first-year physics TAs in a TA professional development course worked in pairs (or sometimes in small groups of three) after completing the MCS individually:

**RQ1.** *What is the role of unguided peer interaction in enhancing TAs' performance in small groups on the MCS?*

**RQ1a.** *What is the frequency with which TAs select the correct answer as a group when one of them answered correctly on their own?*

**RQ1b.** *What is the frequency with which TAs select the correct answer as a group when none of them answered correctly on their own?*

**RQ1c.** *How do the effect sizes vary for questions related to different magnetism concepts on which fewer than 75% of TAs answered correctly individually?*

**RQ2.** *What insights do survey and interview data provide about how peer collaboration impacts the pedagogical skills of TAs, and to what extent they feel confident and prepared to use it in their future teaching?*

## 3. Methodology

*3.1. Participants*

The first-year physics graduate students (pursuing Ph.D.) who participated in this study were from a large public university in the USA and most were TAs for different physics courses (so we call them TAs). They worked individually and then in pairs (with some groups of three) on the MCS [87], which covers introductory magnetism topics. These TAs were enrolled in a graduate-level introduction to teaching course, which is a mandatory teaching related professional development course in the first semester of their first year in the physics department, meeting once a week for 1 h and 50 min. The survey was administered in the early weeks of the semester. As the TAs would take the core graduate-level electricity and magnetism course in their second semester, their knowledge was based on content from previous undergraduate courses.

The survey administration process is shown in the Figure 1 below. The survey data were collected over a two-year period, and for each item, the average score across both years was calculated. TA performance was measured individually and in groups, both prior to and following peer interaction. Initially, 43 TAs completed the MCS individually in approximately 50 min, marking their answers on paper scantrons. They then collaborated with peers—most in pairs, with five groups consisting of three TAs—on the same survey for an additional 50 min, all without professional development course instructor guidance. No feedback was provided to the TAs after the individual portion of the MCS. To better understand the TAs' difficulties with the MCS concepts, 39 students in an upper-level undergraduate electricity and magnetism course were asked to submit written explanations for all MCS questions over the course of two years. We do not consider TAs to be distinctly different from upper-level undergraduates in this context, as they were in the first semester of their graduate program and had not yet taken any graduate-level courses during the initial two weeks of classes; thus, their content knowledge would be similar to that of upper-level undergraduates.

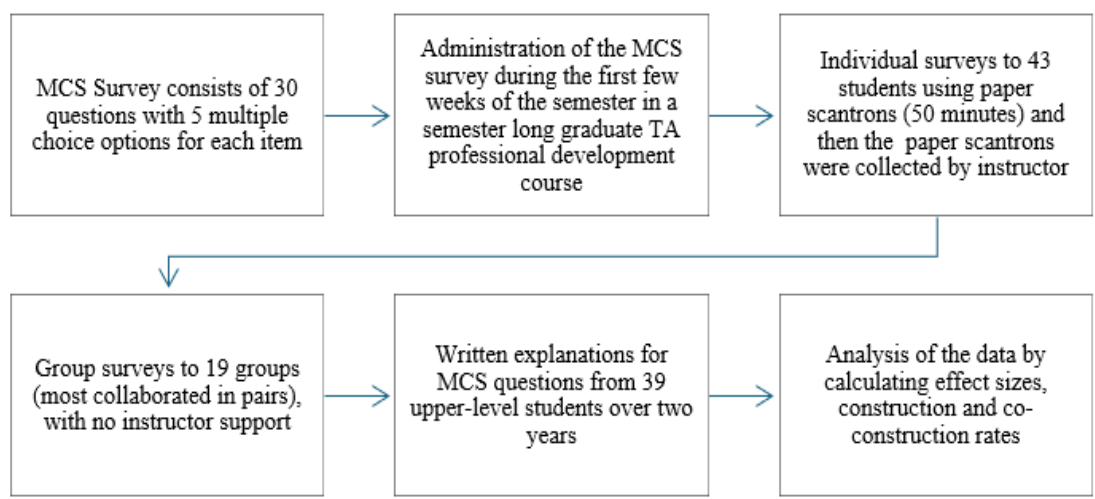

**Figure 1.** A flowchart illustrating the research process of data collection related to collaboration on MCS.

To gain deeper insights into TAs' views on peer collaboration in both TA professional development courses and graduate-level core courses, we conducted interviews with five graduate students who served as physics TAs for introductory courses. We also conducted a written survey in which 18 graduate TAs participated. These TAs were asked about the frequency of peer collaboration in both the TA professional development course and graduate core courses, both inside and outside the classroom. We investigated the nature of their peer interactions, group sizes, the level of participation from all members, and the perceived usefulness of these collaborations. Furthermore, we inquired whether these experiences motivated them to implement peer collaboration techniques in their future teaching, whether they had already applied such methods during recitations as TAs, and if they felt confident using them as instructors in the future. For reference, all survey questions related to peer collaboration are provided in Appendix A. This aspect of the research provided valuable insight into whether the teaching assistants developed the necessary skills to facilitate peer collaboration in their own classrooms.

*3.2. Survey*

The Magnetism Conceptual Survey (MCS) is a validated assessment tool that focuses on topics related to magnetic force and magnetic fields. It is designed to gauge conceptual understanding of magnetism. The survey's final version contains 30 questions, each offering five multiple-choice options. We have done our best to ensure that the context of the questions is clear in our discussion in Appendix B and the two examples in the results section in the main paper, but the readers may find it helpful to refer to a copy of the MCS, which is available through PhysPORT (n.d.) [88].

*3.3. Analysis*

Data analysis involved determining the percentage of TAs who chose the correct answers individually and as part of a group. Knowledge construction for an MCS item is defined as a situation where the group collectively selects the correct answer, but prior to the group discussion, one TA answered correctly, while another answered incorrectly (see Figure 2). The knowledge construction rate was calculated as the percentage of groups that answered correctly in cases where one member initially provided the correct answer and another gave an incorrect one, with this rate being computed for each question. For the groups of three students, the rates of construction and co-construction were calculated using the formula outlined in the referenced study [85].

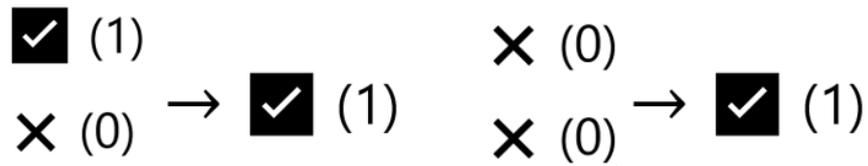

**Figure 2.** Visual Representation for Construction (**left**) and Co-construction (**right**) [84, 85].

Binary scores (0 for incorrect and 1 for correct) were applied to compute the overall scores for both individuals and groups. In this notation, the individual score for a question is listed first, followed by the group score. For instance, in the sequence 011 or 101, the first two digits correspond to the individual responses (incorrect or correct), and the last digit represents the group's response (1 for correct). The rate of construction for each item is calculated using the following formula:

$$\frac{N(10,1) + N(01,1)}{N(10,0) + N(10,1) + N(01,0) + N(01,1)} \times 100\%$$

where $N(10,1)$ denotes the number of groups with one TA answering correctly and the other incorrectly, but the group answering correctly.

Similarly, the co-construction rate refers to the group percentage that correctly answered the question when neither member answered correctly on their own (see Figure 2). In this scenario, the groups receive binary scores of 001, where 0 indicates incorrect individual responses, and 1 represents the correct group answer. The formula for calculating the rate of co-construction is as follows:

$$\frac{N(00,1)}{N(00,0) + N(00,1)} \times 100\%$$

We calculated the effect size using Cohen's $d$ to assess the improvement in performance, comparing individual results to group results [89]. MCS [87] questions were categorized into large-, medium-, and small-effect-size items based on the improvement observed from individual score to group score. An effect size was considered small if Cohen's $d$ was below 0.3, medium if between 0.3 and 0.6, and large if above 0.6 [89]. Cohen's $d$ was calculated using the formula $d = (\bar{X}_1 - \bar{X}_2)/S_{pooled}$, with $\bar{X}_1$ and $\bar{X}_2$ as sample means of individual and group scores and $S_{pooled}$, the pooled standard deviation [89].

## 4. Results and Discussion

**RQ1.** *What is the role of unguided peer interaction in enhancing TAs' performance on the MCS?*

Figures 3 and 4 demonstrate that unguided peer collaboration significantly improved TA performance. In the context of this study, unguided peer collaboration refers to group discussions among TAs that occurred without any guidance from the instructor. Out of 30 MCS questions, 11 had individual correct response rates of 67% or less. After group work, only one question had an average group score below 89%. The average individual score was 74% with a standard deviation of 11%, while the average group score rose to 94% with a standard deviation of 6%. We established a heuristic that an individual score of 75% or higher would indicate good performance, considering potential misreading of questions or answer choices in multiple-choice format.

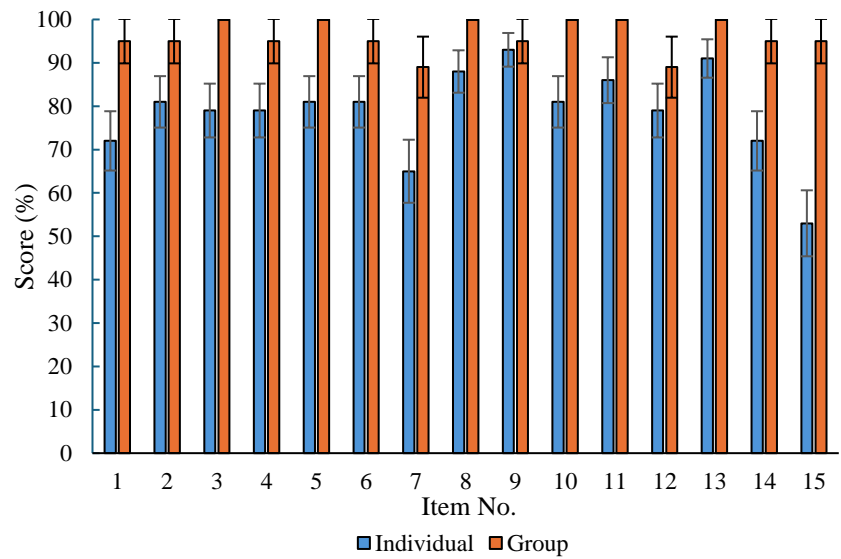

**Figure 3.** Scores for individual and group performance for items 1–15 along with standard errors.

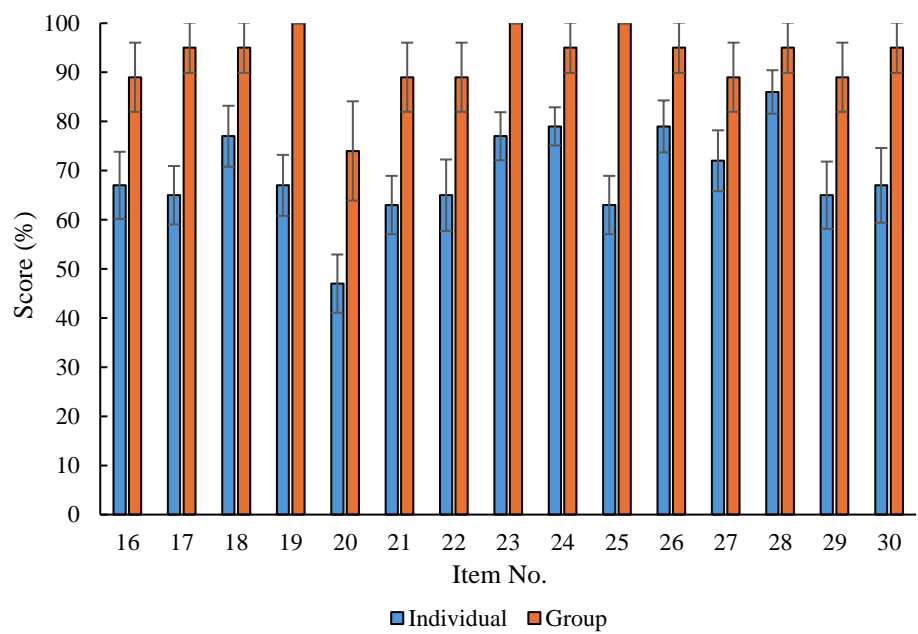

**Figure 4.** Scores for individual and group performance for items 16–30 along with standard errors.

The individual scores varied widely across different MCS questions. Only one item scored below 50% (Q20), and one item fell between 50 and 60% (Q15). Nine items had scores between 60 and 70%, ten between 70 and 80%, seven between 80 and 90%, and two between 90 and 100%. Five items had individual scores greater than 85%, but all items, except Q20, had group scores of 89% or higher following unguided peer collaboration. This strongly suggests that peer interaction, without instructor guidance, is effective in boosting performance on the MCS. It's important to note that TAs did not receive feedback after completing the individual survey. Therefore, comparing individual and group performances for each item offers insights into which concepts TAs were able to grasp on their own and which benefited from peer collaboration, helping to address RQ1c.

Table 1 presents the average individual score and average group score for each question, along with the construction and co-construction rates and effect size [89], showing the improvement in performance from individual to group results. The Construct column indicates the percentage of groups in which at least one TA had the correct answer individually and the group also selected the correct answer during collaboration. A value of 100 in this column means that all such groups arrived at the correct answer as a group. The Co-construct column shows the percentage of groups in which neither TA had the correct answer individually, but the group arrived at the correct answer together. A value of 100 means all such groups co-constructed the correct answer through discussion, while a value of 0 means none did. A dash (—) indicates that there were no groups where both TAs had chosen incorrect answers individually—so co-construction was not applicable. In rare cases (~1% of cases), the group selected an incorrect answer even though both individuals had initially selected the correct one. This occurred in four instances, one of which involved the group leaving the question unanswered, which we counted as incorrect. Table 2 displays the percentage breakdown of TAs choosing the five answer choices for each question individually. It also shows the percentage distribution for groups.

**Table 1.** Individual and group percentage of TAs selecting the correct answer on the MCS [87], construction and co-construction rates for each question, along with the effect size given by Cohen's *d* [89]. The items are listed in descending order of effect size (from individual to group performance).

| Item | Individual | Group | Construct | Co-Construct | Effect Size |
|------|------------|-------|-----------|--------------|-------------|
| 15   | 53         | 95    | 100       | 67           | 0.94        |
| 25   | 63         | 100   | 100       | 100          | 0.91        |
| 19   | 67         | 100   | 100       | 100          | 0.82        |
| 17   | 65         | 95    | 100       | 100          | 0.70        |
| 30   | 67         | 95    | 100       | 0            | 0.66        |
| 23   | 77         | 100   | 100       | -            | 0.65        |
| 3    | 79         | 100   | 100       | 100          | 0.61        |
| 21   | 63         | 89    | 89        | 67           | 0.60        |
| 1    | 72         | 95    | 100       | 100          | 0.57        |
| 14   | 72         | 95    | 90        | 100          | 0.57        |
| 5    | 81         | 100   | 100       | 100          | 0.56        |
| 7    | 65         | 89    | 90        | 0            | 0.56        |
| 10   | 81         | 100   | 100       | 100          | 0.56        |
| 20   | 47         | 74    | 70        | 60           | 0.56        |
| 22   | 65         | 89    | 82        | 100          | 0.56        |
| 29   | 65         | 89    | 91        | 0            | 0.56        |
| 16   | 67         | 89    | 75        | 100          | 0.51        |
| 11 | 86 | 100 | 100 | - | 0.48 |
| 18 | 77 | 95 | 89 | - | 0.47 |
| 4 | 79 | 95 | 100 | 50 | 0.43 |
| 8 | 88 | 100 | 100 | - | 0.43 |
| 24 | 79 | 95 | 88 | - | 0.43 |
| 26 | 79 | 95 | 100 | - | 0.43 |
| 27 | 72 | 89 | 78 | 100 | 0.42 |
| 2 | 81 | 95 | 100 | 0 | 0.38 |
| 6 | 81 | 95 | 83 | 100 | 0.38 |
| 13 | 91 | 100 | 100 | - | 0.38 |
| 12 | 79 | 89 | 80 | 0 | 0.27 |
| 28 | 86 | 95 | 100 | - | 0.27 |
| 9 | 93 | 95 | 100 | - | 0.07 |

**Table 2.** Percentage of individuals and groups selecting each answer option for every item. The correct answer is **bold**. TAs who skipped (S) a question are marked as incorrect for that item. ($N_I$ = 43 for individuals and $N_G$ = 19 for groups).

| Item # | Group/Individual | A | B | C | D | E | S | Construction | Co-Construction |
|---|---|---|---|---|---|---|---|---|---|
| 1 | Individual | 5 | 0 | 5 | **72** | 19 | 0 | 100 | 100 |
|   | Group | 0 | 0 | 0 | **95** | 5 | 0 | | |
| 2 | Individual | 7 | 9 | **81** | 0 | 0 | 3 | 100 | 0 |
|   | Group | 0 | 5 | **95** | 0 | 0 | 0 | | |
| 3 | Individual | 9 | 9 | **79** | 0 | 0 | 3 | 100 | 100 |
|   | Group | 0 | 0 | **100** | 0 | 0 | 0 | | |
| 4 | Individual | **79** | 0 | 2 | 7 | 9 | 3 | 100 | 50 |
|   | Group | **95** | 0 | 0 | 0 | 0 | 5 | | |
| 5 | Individual | **81** | 5 | 0 | 2 | 12 | 0 | 100 | 100 |
|   | Group | **100** | 0 | 0 | 0 | 0 | 0 | | |
| 6 | Individual | 2 | **81** | 0 | 5 | 7 | 5 | 83 | 100 |
|   | Group | 0 | **95** | 5 | 0 | 0 | 0 | | |
| 7 | Individual | 0 | 2 | 14 | **65** | 19 | 0 | 90 | 0 |
|   | Group | 0 | 0 | 5 | **89** | 5 | 0 | | |
| 8 | Individual | 2 | 0 | 2 | **88** | 5 | 3 | 100 | - |
|   | Group | 0 | 0 | 0 | **100** | 0 | 0 | | |
| 9 | Individual | 0 | **93** | 2 | 5 | 0 | 0 | 100 | - |
|   | Group | 0 | **95** | 0 | 0 | 0 | 5 | | |
| 10 | Individual | 2 | 0 | 0 | **81** | 16 | 0 | 100 | 100 |
|    | Group | 0 | 0 | 0 | **100** | 0 | 0 | | |
| 11 | Individual | 7 | 0 | **86** | 5 | 2 | 0 | 100 | - |
|    | Group | 0 | 0 | **100** | 0 | 0 | 0 | | |
| 12 | Individual | 0 | 7 | 9 | **79** | 2 | 3 | 80 | 0 |
|    | Group | 0 | 0 | 11 | **89** | 0 | 0 | | |
| 13 | Individual | **91** | 2 | 5 | 0 | 0 | 3 | 100 | - |
|    | Group | **100** | 0 | 0 | 0 | 0 | 0 | | |
| 14 | Individual | 0 | 21 | 0 | 5 | **72** | 3 | 90 | 100 |
|    | Group | 0 | 5 | 0 | 0 | **95** | 0 | | |
| 15 | Individual | 14 | 12 | 12 | 5 | **53** | 5 | 100 | 67 |
|    | Group | 5 | 0 | 0 | 0 | **95** | 0 | | |
| 16 | Individual | **67** | 9 | 23 | 0 | 0 | 0 | 75 | 100 |
|    | Group | **89** | 5 | 5 | 0 | 0 | 0 | | |
| 17 | Individual | 14 | 2 | 12 | **65** | 5 | 3 | 100 | 100 |
| | | Group | 0 | 0 | 5 | **95** | 0 | 0 | |
| 18 | Individual | 0 | 12 | 7 | 0 | **77** | 5 | 89 | - |
| | Group | 0 | 0 | 5 | 0 | **95** | 0 | | |
| 19 | Individual | 9 | 5 | 2 | 14 | **67** | 3 | 100 | 100 |
| | Group | 0 | 0 | 0 | 0 | **100** | 0 | | |
| 20 | Individual | 7 | **47** | 9 | 19 | 16 | 3 | 70 | 60 |
| | Group | 0 | **74** | 0 | 16 | 11 | 0 | | |
| 21 | Individual | 9 | 23 | **63** | 2 | 0 | 3 | 89 | 67 |
| | Group | 0 | 11 | **89** | 0 | 0 | 0 | | |
| 22 | Individual | 2 | 2 | **65** | 28 | 0 | 3 | 82 | 100 |
| | Group | 0 | 0 | **89** | 11 | 0 | 0 | | |
| 23 | Individual | 0 | 2 | **77** | 2 | 16 | 3 | 100 | - |
| | Group | 0 | 0 | **100** | 0 | 0 | 0 | | |
| 24 | Individual | 0 | **79** | 12 | 2 | 5 | 3 | 88 | - |
| | Group | 0 | **95** | 0 | 0 | 5 | 0 | | |
| 25 | Individual | 5 | 7 | 21 | 2 | **63** | 3 | 100 | 100 |
| | Group | 0 | 0 | 0 | 0 | **100** | 0 | | |
| 26 | Individual | 2 | 7 | 2 | 7 | **79** | 3 | 100 | - |
| | Group | 5 | 0 | 0 | 0 | **95** | 0 | | |
| 27 | Individual | **72** | 5 | 2 | 2 | 14 | 5 | 78 | 100 |
| | Group | **89** | 11 | 0 | 0 | 0 | 0 | | |
| 28 | Individual | **86** | 7 | 2 | 0 | 2 | 3 | 100 | - |
| | Group | **95** | 5 | 0 | 0 | 0 | 0 | | |
| 29 | Individual | 14 | **65** | 5 | 0 | 14 | 3 | 91 | 0 |
| | Group | 0 | **89** | 0 | 0 | 5 | 5 | | |
| 30 | Individual | 19 | **67** | 2 | 2 | 7 | 3 | 100 | 0 |
| | Group | 0 | **95** | 0 | 5 | 0 | 0 | | |
| | | AVERAGE RATES | | | | | | **93** | **72** |

**RQ1a.** *What is the frequency with which TAs select the correct answer as a group when one of them answered correctly on their own?*

We found that construction occurred in 93% of eligible cases across all MCS questions, indicating that when one TA knew the correct answer, the group response was correct. Tables 1 and 2 provide a detailed breakdown of construction for each item. These results highlight the high effectiveness of peer collaboration.

**RQ1b.** *What is the frequency with which TAs select the correct answer as a group when none of them answered correctly on their own?*

We found that co-construction occurred in 72% of eligible cases across all questions, meaning that when none of the TAs knew the correct answer, the group response was correct. Tables 1 and 2 provide detailed information on co-construction for each item. These findings also highlight the effectiveness of peer collaboration.

**RQ1c.** *How do the effect sizes vary for questions related to different magnetism concepts on which fewer than 75% of TAs answered correctly individually?*

Here, we present two representative examples of questions for which peer collaboration led to large improvements from individual to group scores. Additional examples of items with large, medium or small effect sizes are provided in the Appendix B [89]. These findings can help instructors understand the amount of improvement on various MCS concepts based upon collaboration with peers and compare it with improvement in their own courses if similar peer collaboration is used in their classes.

Q15 (Figure 5) had the highest improvement among all MCS questions. This question assesses understanding of the net force direction on a positively charged particle moving through crossed uniform electric and magnetic fields. In Figure 5, the particle enters the fields with speed v, where the magnetic field points to the left, and the electric field points into the page. Initially, the individual score was the second lowest at 53%, but it rose significantly to a group score of 95%, as shown in Table 1. The effect size for this question was an impressive 0.94. Additionally, this item showed 100% construction and 67% co-construction rates. Table 2 reveals that the most selected individual answer was option e ("Not enough information to determine the direction of the net force"), which is correct. However, no single incorrect answer dominated; responses were distributed across options, with 14% selecting option a (to the left), 12% selecting option b (into the page), and another 12% selecting option c (out of the page). This variety of incorrect answers likely prompted rich discussions among TAs, contributing to strong group performance. TAs who selected option b often misapplied the right-hand rule, incorrectly concluding that the magnetic force directed into the page and combining it with the electric force in the same direction. Meanwhile, TAs choosing options a or c appeared to struggle with determining the net force from the two contributing forces and may have chosen randomly. An upper-level student choosing option a explained, "the electric field pushes it left and the magnetic field pushes it out by the RHR [right hand rule]". Prior research [87] indicates that both introductory and upper-level students faced significant difficulties with this question, showing minimal improvement in performance even after instruction. However, the remarkable effectiveness of peer interaction in enabling TAs to converge on the correct response for Q15 underscores the substantial benefits of providing opportunities for peer collaboration.

A positively charged particle moving upward in the plane of the paper with a speed $v$ enters a region with uniform magnetic field of magnitude $B$ directed to the left (⟵) and a uniform electric field of magnitude $E$ directed into the page (⊗).

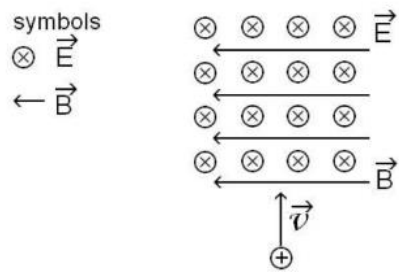

What is the direction of the net force on the particle due to the fields at the instant it enters?

(a) To the left (⟵)
(b) Into the page (⊗)
(c) Out of the page (⊙)
(d) No net force
(e) Not enough information to determine the direction of the net force.

**Figure 5.** MCS item 15.

Q25 was the item with the second highest improvement on the MCS. The problem involved two parallel wires (wire 1 and wire 2), each carrying current to the right (east), placed in an external magnetic field directed out of the page. TAs had to determine the net magnetic force on wire 2, which was positioned below wire 1. The individual score of

63% increased to 100% after group work, as shown in Table 1. It had an effect size of 0.91 and a gain of 37%. This item had 100% rates of both construction and co-construction. Table 2 reveals that 21% of TAs chose option c (away from wire 1), 7% chose option b (toward wire 1), and 5% chose option a (out of the page). TAs who selected option c had only considered the force from the external magnetic field and neglected the force from wire 1's magnetic field. For instance, one upper-level student chose option c and stated, "Because of the right-hand rule, the field the wire is [in], regardless of wire one will push the wire in the downward direction." However, after discussing with their peers, all TA groups recognized their mistake and took both forces into account to determine the correct direction of the net force on wire 2 (option e). According to the validation paper on MCS [87], Q25 was the most challenging question on the test. This demonstrates that TAs can greatly benefit from unguided group work, even when tackling the most challenging concepts in magnetism.

**RQ2.** *What insights do survey and interview data provide about how peer collaboration impacts the pedagogical skills of TAs, and to what extent they feel confident and prepared to use it in their future teaching?*

Based on the survey we conducted among graduate students who served as TAs in the physics department, we found that a majority frequently participated in group work. Specifically, 55% reported engaging in group activities either in every class or a few times per month during the TA professional development course, as shown in Figure 6a. A noticeable contrast emerged when comparing group work inside and outside the classroom for graduate courses. TAs appeared to collaborate more extensively outside the classroom in their graduate-level courses, as illustrated in Figure 6b and 6c. However, only about one-third of the TAs reported collaborating with others on a conceptual physics survey, while the rest either did not or were unsure, as seen in Figure 6d. We note that Figure 6e shows that 50% of the TAs expected the same or a higher score individually on a conceptual survey after working with peers compared to their group score on the same survey. We wanted to explore why 50% of the students expected to score lower individually after participating in group work, so we addressed this in the interview. One TA explained, "I don't know if I would remember everything that me and the group discussed when I'm taking it on my own, but I think I would understand things better, so probably the same or lower." This type of comment suggests that it is crucial to foster both positive interdependence and individual accountability during peer collaboration, ensuring that each member actively contributes, supports one another, and benefits maximally from the group work.

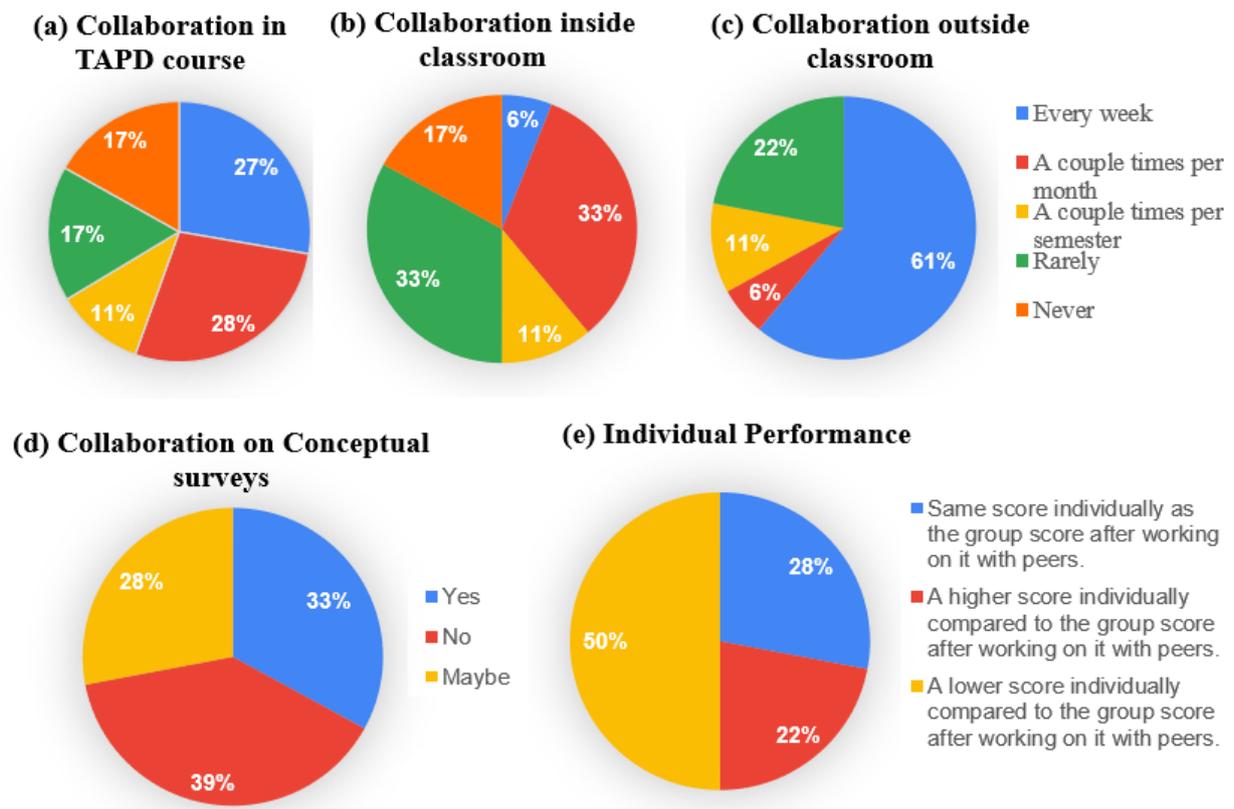

**Figure 6.** Responses in percentage from TAs about collaboration in (**a**) Teaching Assistant Professional Development (TAPD) course, (**b**) Graduate courses inside the classroom, (**c**) Graduate courses outside the classroom, (**d**) Physics Conceptual surveys, and (**e**) Prediction about their individual performance after group work on a physics conceptual survey (Read clockwise from the top, beginning with the blue segment).

We asked TAs about their experiences collaborating with peers in their graduate-level core courses. The TAs could select multiple courses and group sizes in which they collaborated, as these could have varied across courses, and the percentages are reported accordingly as seen in Figure 7. Regarding the courses in which they worked with others, 78% reported collaborating in Electricity and Magnetism, followed by 56% in Statistical Mechanics and Thermodynamics, and 50% in Dynamical Systems. In terms of group size, 78% indicated they typically worked in groups of 2–3 students. When asked whether all group members contributed equally during collaboration, 65% said yes. As for the perceived usefulness of these collaborations, 59% found them very useful, while only 11% considered them not useful. Finally, we asked whether participating in group work—either in the TA Professional Development (TAPD) course or in other graduate courses—inspired them to incorporate collaborative learning strategies in their own teaching, whether as a TA or future instructor. While responses were somewhat mixed, a greater percentage of TAs reported being motivated to apply collaborative techniques in their own classrooms.

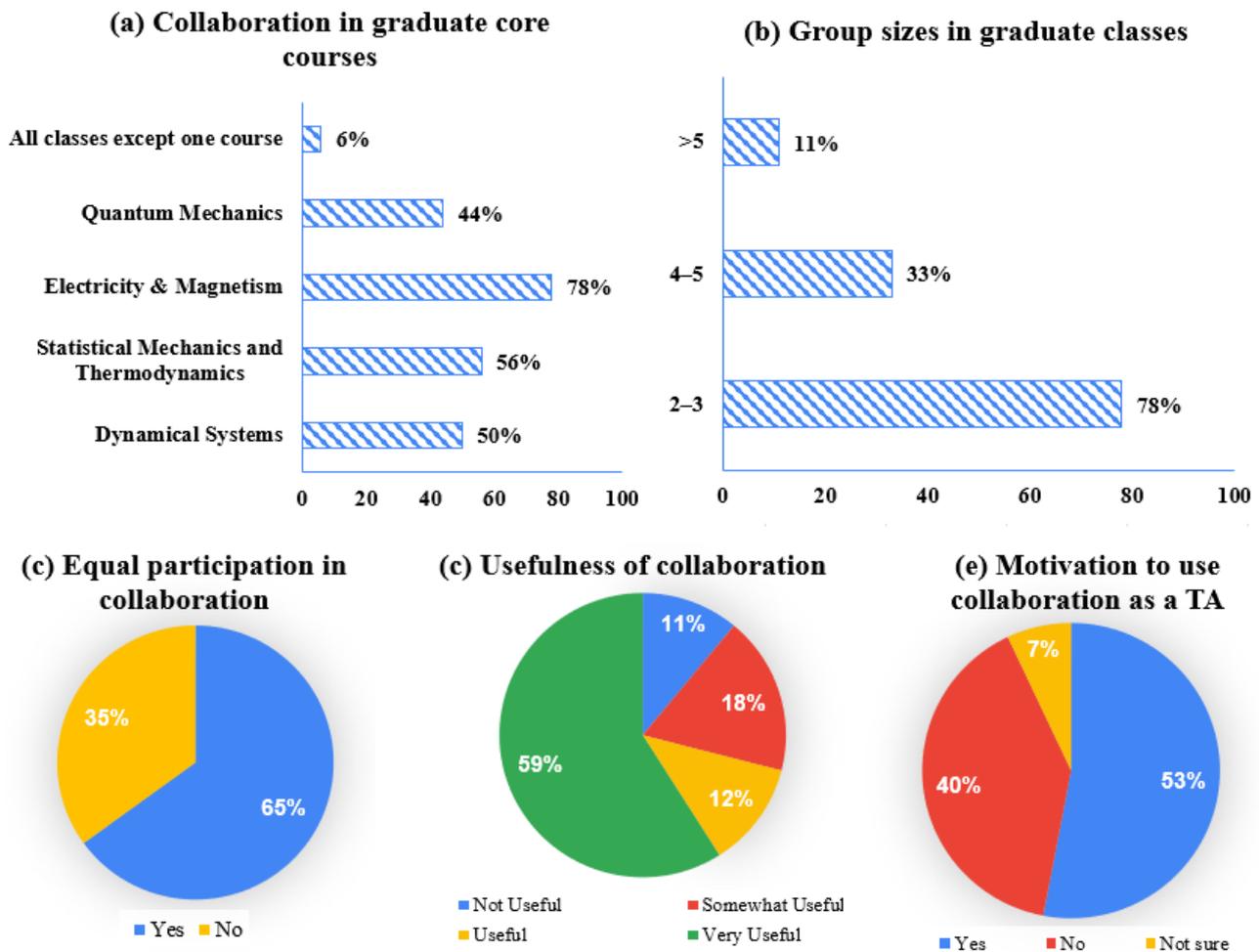

**Figure 7.** Responses in percentage from TAs about (**a**) the graduate core courses in which the TAs collaborated with each other, (**b**) the group sizes in graduate classes, (**c**) if there was equal participation from all group members during collaboration, (**d**) usefulness of collaboration, and (**e**) if participating in group work with other TAs in a TAPD course or in other graduate courses motivated them to use collaborative learning techniques in their own teaching either as a TA or an instructor (Read clockwise from the top, beginning with the blue segment).

We also asked the TAs about their teaching experiences, specifically regarding the use of collaborative learning strategies as seen in Figure 8a–f. When asked whether they had implemented collaborative learning techniques in the classes they taught, over two-thirds responded positively. Similarly, more than two-thirds reported allowing students to form their own groups. Regarding group size, 80% indicated that students typically worked in groups of 2–3, while 20% mentioned groups of 4–5 students. As for whether group work was tied to a grade incentive, responses were evenly split. Further, 71% of the TAs said they plan to regularly incorporate collaborative learning in their future teaching and an additional 23% said they would do so occasionally, demonstrating strong enthusiasm for peer collaboration. When asked about their confidence in facilitating collaborative learning, only 7% reported feeling not confident, while the remaining TAs ranged from somewhat confident to very confident. Overall, the responses were highly positive, indicating that most TAs not only value peer collaboration but also feel equipped to implement it effectively in their teaching based upon their prior experiences.

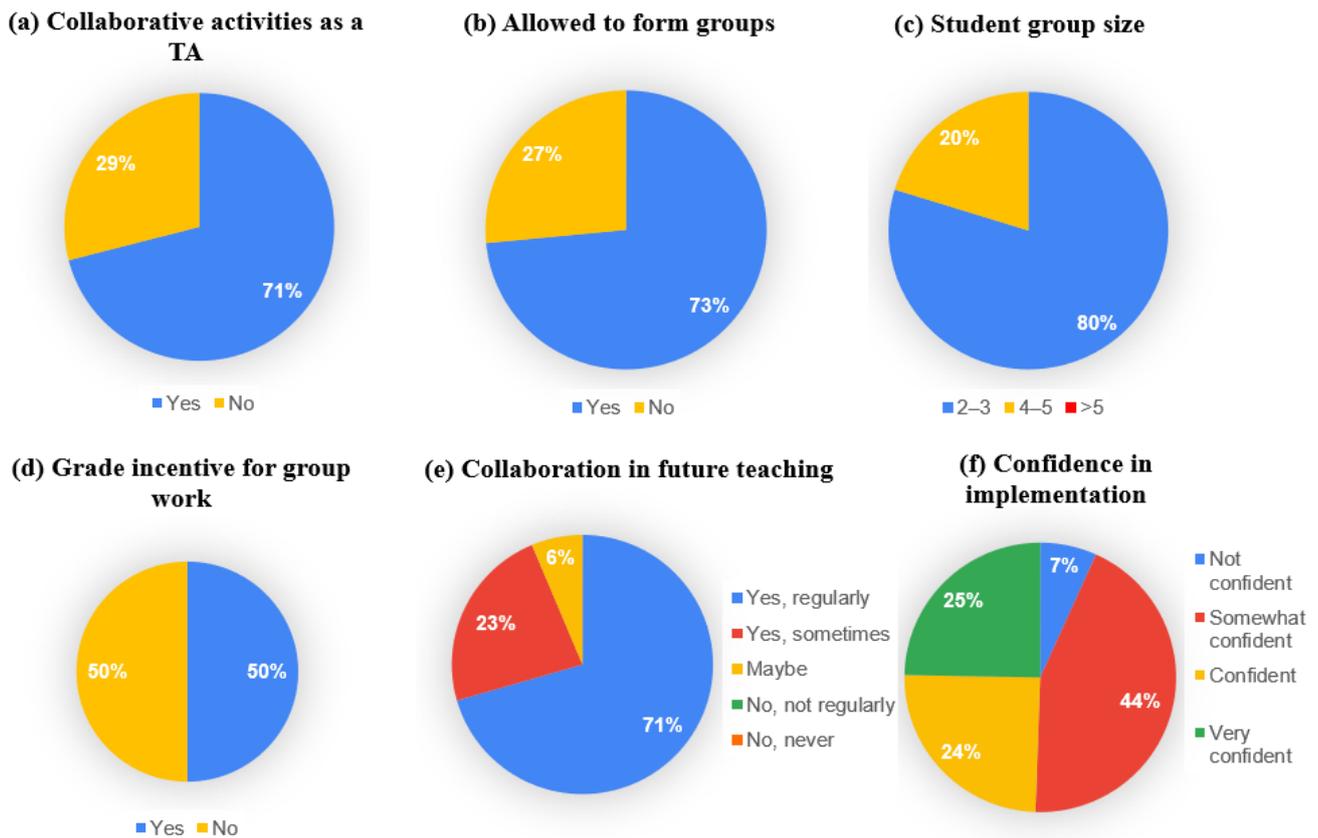

**Figure 8.** Percentage of TAs in response to whether they (**a**) applied collaborative learning techniques in the classes they taught, (**b**) allowed their students to form groups, (**c**) the group sizes in which their students collaborated, (**d**) if there was a grade incentive for group work, (**e**) if they plan to use collaborative learning techniques in future teaching and (**f**) their confidence in facilitating collaborative learning techniques among their students (Read clockwise from the top, beginning with the blue segment).

We were also interested in understanding the group interactions in graduate-level courses. Some TAs shared that they collaborated on homework problems, often meeting outside the classroom in small groups. One TA described the experience: "You first needed to find people who would work well together meaning they were supportive and open to others' ideas. Then we would find an empty room and show each other what we did to try to solve the problems, then collaborate at the board, trying to solve the rest". Another TA emphasized the importance of following up group work with individual effort, stating, "I would work on problem sets by myself, and if I got stuck, I would work with my classmates. Once the issue is resolved, I would usually work by myself to finish the assignment." These reflections illustrate the range of experiences TAs had with collaboration, highlighting the importance of group work in graduate classes.

TAs shared a range of experiences regarding the implementation of collaborative learning techniques in their teaching. Many mentioned incorporating group work during recitations, quizzes, or class discussions. Several expressed that collaborative learning was helpful for students, with one TA stating "…I found that this was helpful in exposing students to multiple viewpoints for tackling the problems, leading to an increase in successful problem solving". Some commented that students seemed to appreciate these opportunities and often performed better when allowed to collaborate. One TA shared their experience administering in-class quizzes and noted, "When I was a TA, I administered in class quizzes…I found that this usually helped students arrive at the right answer more often." However, instructor constraints limited a couple of TAs from applying group work strategies more broadly. One TA felt it worked well for short, in-class discussions but not for longer, take-home group projects.

TAs provided a range of examples regarding how peer collaboration has influenced their teaching practices. Many noted that working in groups helped them gain confidence, have a sense of belonging, and understand that others often share similar struggles. One TA shared their experience, "There were elements of community formation that came with group work in that course, which reminded me of how important it can be for students to interact to feel safe in a learning environment." Another TA stated, "Peer collaboration makes you understand that it is not only you who finds a problem challenging, and other people usually have the same questions as you do. So, it helps boost your confidence". Several emphasized the importance of fostering an environment where students can co-construct knowledge, benefit from different perspectives, and engage in mutual problem solving. Some TAs shared that they actively incorporate peer collaboration by structuring activities with conceptual and calculation-based questions, encouraging discussion during class, and providing space for collaborative group work. A few mentioned that their own positive experiences with peer collaboration inspired them to adopt similar strategies in their teaching, sharing that the, "…best part of it [peer collaboration] was to learn that problems can be solved discussing it with at least one more person, because there's always one thing that we don't see right away that the other person or people might see it first, and vice versa. So doing it myself and also encouraging it in my class [that I teach] to my students is and has been very important. And I will keep practicing it when possible".

As for improving the TA professional development course, many suggested incorporating more practical elements—such as sample lessons, real-class scenarios, and peer-led demonstrations—to build confidence in facilitating collaborative learning. Suggestions included rotating group members, observing videos of effective facilitation, and discussing how to manage group dynamics, especially in larger classes or when students are disengaged. Overall, the responses reflected thoughtful insights into the value of peer collaboration and a desire for more support and concrete strategies in applying it effectively in the classroom.

## 5. Conclusions and Instructional Implications

This study investigated the extent to which TAs in a mandatory TA professional development course (designed for these early educators who were first-year physics Ph.D. students) improved their content knowledge when collaborating with their peers while they also gained pedagogical knowledge about collaborative learning. Investigating the improvement in their performance when working individually and after working with peers is invaluable for effective instruction because collaboration with peers can improve both their content and pedagogical knowledge. We found that the TAs who completed the MCS individually showed a significant improvement in their performance after working with their peers without any guidance from the TA professional development course instructor. The fact that they showed a significant improvement not only via construction but also co-construction of knowledge after peer collaboration indicates that the peer collaboration was productive. In particular, they were able to articulate their thought processes while working through MCS problems and help each other. This type of unguided peer interaction can be an important component of the professional development of early educators such as TAs and can help them support their teaching both via improved content and pedagogical knowledge.

TAs performed extremely well on all MCS questions in groups, with a group score ≥ 89%, except Q20. Q20 was the item with the lowest group score of 74%, pertaining to the movement of a charged particle in a magnetic field. We analyzed the questions by considering the effect sizes from individual to group performance. We find that most of these questions had medium or high effect sizes in terms of improvement from individual to group work. We also found that construction and co-construction happened very frequently. The average construction rate was 93% and the average co-construction rate was observed to be 72% across all the questions. In other words, when only one of the TAs in the group was correct, the group gave the correct answer 93% of the time, and when both TAs were wrong individually, they provided the correct answer 72% of the time. There were a few questions which had a low rate of construction but showed a significant improvement in their group performance due to co-construction. This suggests that most of the MCS items come under the zone of proximal facilitation of the group. Overall, the findings from the MCS underscore the effectiveness of peer collaboration in improving understanding of the complex physics concepts while helping TAs learn collaborative learning pedagogy via first-hand experience with it. The TAs, many of whom struggled with difficult magnetism concepts such as the right-hand rule and magnetic force direction, saw significant improvements in their performance after working with peers. This highlights the power of these early educators learning collaboratively while internalizing the value of collaborative learning pedagogy, even in the absence of the TA professional development course instructor's feedback during their collaboration.

The survey conducted among graduate students who served as TAs in the physics department revealed that a majority actively engaged in group work, particularly outside the classroom, in graduate-level courses. Our findings indicate that approximately two-thirds of the TAs reported collaborating with peers either weekly or a few times per semester (Figure 6a). Nearly 90% of TAs found peer collaboration to be useful (Figure 7d), and about 95% expressed intentions to incorporate collaborative learning techniques in their future teaching (Figure 8e). These findings strongly suggest that many TAs not only valued the experience of peer collaboration but also developed an appreciation for its pedagogical utility. Their expressed intention to apply these strategies as future instructors, points to meaningful pedagogical development showing that the experience supported not only content learning but also the development of instructional skills. In particular, TAs overwhelmingly expressed positive views on collaborative learning, with the majority implementing group work strategies in their teaching. They valued peer collaboration for boosting confidence and enhancing problem-solving skills, although they also noted challenges such as unequal participation in longer-term group projects.

The findings underscore the importance of fostering both positive interdependence and individual accountability in peer collaboration. While TAs found collaboration helpful in their learning and teaching experiences, they also highlighted the need for practical strategies to improve group dynamics and ensure balanced participation. Many suggested that the TA professional development course could benefit from more hands-on activities, such as peer-led demonstrations and strategies for managing group work in larger or disengaged classes. Overall, the survey indicates strong enthusiasm for collaborative learning among TAs, along with a desire for better tools to facilitate effective peer collaboration in the classroom.

Our findings are consistent with prior research using the Conceptual Survey of Electricity and Magnetism (CSEM) [84], which showed that unguided peer collaboration among physics TAs led, e.g., to similarly high rates of construction, where groups arrived at correct answers when only one member initially had the correct response. Importantly, these collaborative experiences supported the development of pedagogical skills as TAs engaged in diagnosing each other's difficulties, articulating reasoning, and negotiating instructional approaches which are key elements of effective teaching practice. The parallels between the MCS and CSEM contexts reinforce the robustness of unguided collaboration as a valuable tool for both conceptual and instructional development.

However, we hypothesize that focusing on the peer collaboration among first-year graduate TAs for the MCS in this study may have positively influenced peer collaboration outcomes, as they are likely to have a close-knit community, being part of the same cohort. It would be valuable to investigate how peer collaboration is affected when individuals in a group lack a shared sense of belonging [90, 91], which is critical for productive communities of practice (CoPs). Since professional development leaders have very limited time, organizing and supporting effective CoPs involving early STEM educators in college (such as in the case of the TAs in this study) or pre-college educators is central to helping them tap into each other's strengths. The shared goals of the educators in the CoP and their sense of belonging in the community can help them collaborate effectively to improve their content knowledge while also helping them learn valuable pedagogical approaches that they can employ in their own classes with their students.

Thus, by creating and supporting a CoP, the dual benefits of improving content and pedagogical knowledge of early college educators (such as the TAs) discussed here can be adapted and applied to teacher professional development programs, including those for early pre-college STEM educators who may not have an in-depth knowledge of the subject or the pedagogy. Just as the TAs benefitted from peer collaboration on the MCS, pre-college educators can enhance their understanding of STEM concepts while working collaboratively with their peers and also improve their pedagogical knowledge of collaborative learning. We note that in the US, less than 50% of the high school physics teachers have a major or minor degree in physics, which can result in gaps in many early educators' understanding of core concepts they are supposed to teach. Through the types of interactions discussed here in a supportive CoP, teachers can share their knowledge, challenge their assumptions, and deepen their understanding of difficult concepts. As noted, this process can help strengthen their content knowledge and can also help them build important pedagogical skills. As they engage in discussions and problem-solving with peers, they can learn the benefits of collaborative learning pedagogy and how it can support their students to think critically, approach complex concepts, and articulate concepts effectively. Thus, positive collaborative experience in a CoP can be transferred to the early educators' teaching practices, and they can become inspired to help their students work together in groups, share knowledge, and support each other's learning.

In summary, the research described here for physics TAs can be adapted and be particularly valuable for early educators across disciplines, who are supported to be part of a mutually beneficial CoP of like-minded educators. It can help incentivize and motivate them to create a classroom environment where peer-to-peer learning is encouraged, helping their students develop both their conceptual understanding and collaboration skills. In this way, peer collaboration can have a two-fold benefit for educators, i.e., it can enhance their content knowledge and also equip them with the tools to foster the same collaborative mindset in helping their own students learn. Since the development of content and pedagogical knowledge discussed in this research does not require time investment from professional development leaders, early career educators can be supported via opportunities to participate in appropriate CoPs and engage with one another to collectively enhance their content and pedagogical knowledge. They can connect in these CoPs either in person within a specific geographic region or virtually (e.g., via Zoom) and participate in collaborative activities that have the potential to significantly strengthen their content knowledge and pedagogical approaches, e.g., pertaining to collaborative learning discussed here.

## Appendix A. Survey on Peer Collaboration

*Appendix A.1. Participation in Peer Collaboration*

1. How frequently did you engage in group work with other graduate students in the TA professional development course?
    1. Every class
    2. A couple times per month
    3. A couple times per semester
    4. Rarely
    5. Never
2. Have you ever participated in group work specifically while working on the Magnetism Conceptual Survey (MCS) or any other conceptual surveys in the same TA professional development course?
    1. Yes
    2. No
    3. Maybe
3. Suppose you took a physics conceptual survey three times in the following order: individually, with a peer in a group, individually. Would you obtain the same score individually as your group score on the same multiple-choice survey after working on it with peers?
    1. I would usually expect to obtain the same score individually as the group score after working on it with peers.
    2. I would usually expect to obtain a higher score individually compared to the group score after working on it with peers.
    3. I would usually expect to obtain a lower score individually compared to the group score after working on it with peers.
4. How frequently did you engage in group work with other graduate students in any other graduate level physics courses inside the classroom?
    1. Every class
    2. A couple times per month
    3. A couple times per semester
    4. Rarely
    5. Never
5. How frequently did you engage in group work with other graduate students in any other graduate level physics courses outside the classroom?
    1. Every week
    2. A couple times per month
    3. A couple times per semester
    4. Rarely
    5. Never
6. If you did engage in group work, which course was it?
    - Dynamical Systems
    - Statistical Mechanics and Thermodynamics
    - Electricity & Magnetism
    - Quantum Mechanics
    - Other…………
7. What was the group interaction like in the graduate level courses?
8. How large was the group?
    - 2–3
    - 4–5
    - >5

9. Did all group members equally participate?
    1. Yes
    2. No
10. How useful did you find the collaborative activities?
    1. Not useful
    2. Somewhat useful
    3. Useful
    4. Very useful

*Appendix A.2. Pedagogical Knowledge and Practices*

11. Did participating in group work with other TAs in a TA professional development course or graduate students in other courses motivate you to use collaborative learning techniques in your own teaching? Explain.
12. Have you applied collaborative learning techniques in the classes you taught so far? If so, what was your experience like implementing these techniques? Explain.
13. How large were the groups in which students participated in peer collaboration in the classes you have taught?
    - 2–3
    - 4–5
    - >5
14. Were the students allowed to form their own groups? Explain.
15. What types of problems did the students work on?
16. Was there any grade incentive associated with the group work?
    - Yes
    - No
17. Please share an example of how peer collaboration influenced your teaching practices as a TA or as an independent instructor.
18. Would you consider using collaborative activities with your own students in future teaching as an instructor?
    - Yes, regularly
    - Yes, sometimes
    - Unsure
    - No, not regularly
    - No, never
19. How confident are you in facilitating collaborative learning techniques among your students?
    - Not Confident
    - Somewhat Confident
    - Confident
    - Very Confident
20. What can be done in the TA professional development course to increase your confidence in facilitating collaborative learning techniques?
21. Do you have any suggestions for how to better evaluate the impact of collaborative learning on TAs in a TA professional development course?

## Appendix B. Large, Medium and Small Effect Size Questions

*B.1 Large Effect Size*

There were seven Magnetism Conceptual Test (MCS) questions with large effect size ($d > 0.6$) [88]. Here we analyze those items to illustrate the features that led to large effect size for the concepts covered in these questions. We have not discussed Q3 and Q23 with large effect sizes because the individual scores were greater than 75% for these items.

Q19 is very similar to Q15 (discussed in the main text) with only difference in the direction of electric field pointing downwards, magnetic field into the page and the charged particle moving to the right. This item demonstrated the third-largest improvement, increasing from 67% in individual performance to 100% in group performance. From Table I in the main text, we can see that the effect size is 0.82 and the gain is 33% for this item. The high group score was a result of 100% construction and co-construction rates. The most common individual incorrect choice was option d (no net force) followed by option a (upwards) as shown in Table II in the main text. The TAs who chose option d correctly figured out that the directions of forces due to the electric and magnetic fields are opposite to each other, but they did not realize that they needed more information on their relative strength to figure out the direction of the net force.

The figure below shows two situations in which a positively charged particle moving with speed $v$ through a uniform magnetic field of magnitude $B$ experiences a magnetic force $\vec{F}_B$ due to the field.

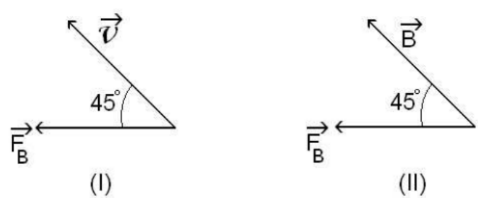

Which orientations are physically reasonable?
(a) I
(b) II
(c) Both I and II.
(d) Neither of them.
(e) Not enough information to decide.

Fig B1.1: MCS item 17

Q17 (Fig. S1) addresses the direction of the magnetic force ($\vec{F}$) on a positively charged particle relative to its velocity ($\vec{v}$) and magnetic field ($\vec{B}$). In this question, TAs are presented with two diagrams and asked whether these orientations are physically plausible. They must recognize that the magnetic force is always perpendicular to both the velocity and the magnetic field. Table I in the main text shows that Q17 had an individual score of 65% and after peer interaction, it increased to 95% with a high effect size of 0.70 and a gain of 30%. From Table II in the main text, we see that 14% of TAs chose option a (I), and 12% selected option c (I and II), suggesting difficulty in determining the direction of the magnetic force. Some TAs seemed to struggle with the concept that the force must be perpendicular to both the velocity and magnetic field, as indicated by the cross product. The written explanations provided by upper-level students who chose option a revealed that many correctly understood that the magnetic force is perpendicular to the magnetic field but mistakenly thought that the velocity and force could be at any angle to each other. Those who chose option c had an incorrect notion that both the orientations are possible using the right-hand rule. One upper-level student who chose option a stated, "The magnetic force has to be perpendicular to magnetic field so 2 isn't possible. 1 seems reasonable." and another who chose option c stated, "I am unsure of what this problem is labeling but given the right-hand rule we know that if the $F_b$ [magnetic force] has a perpendicular component in the right direction determined by the right-hand rule then it is possible to happen. It is very hard to tell which way the arrows of the given forces are pointing."

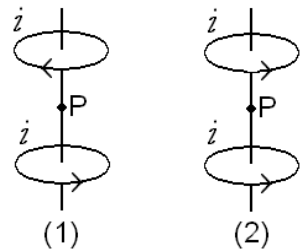

Fig B1.2: MCS item 30.

Q30 compares the strength of the magnetic fields at the midpoint of two current carrying circular loops as shown in Fig. S2. Table I in the main text shows that Q30 had an individual score of 67% and it increased to a group score of 95% resulting in an effect size of 0.66. Table II in the main text shows that 19% of the TAs individually thought that the magnitude of the magnetic field is greater in case 1. Some TAs incorrectly concluded that the magnetic fields due to the two current loops get added up in case 1 using right-hand rule.

*B.2. Medium Effect Size*

Table I in the main text shows that 20 out of 30 items had a medium effect size ($0.3 \leq d \leq 0.6$) [88]. It is noteworthy that each of these items had an average individual score below 75%. Most of these questions did not have a dominant incorrect answer choice but rather a mix of incorrect answers. We hypothesize that this mix of answers may have partly led to a fruitful discussion and improved group performance. Analysis of items 1, 7, 14, 16, 20, 21, 22, 27 and 29 shows several characteristics that may lead to medium effect size on these items.

Q21 investigates understanding of the right-hand rule in a scenario where a very long straight wire carries current to the right and is placed in a uniform magnetic field pointing out of the page. The TAs were tasked with determining the direction of the magnetic force on the wire, which is downward with respect to the paper. This question had an individual score of 63%, which improved to 89% after peer collaboration, corresponding to an effect size of 0.60 (Table I). Table II in the main text reveals that 23% of TAs initially chose option b (upward), and 9% selected option a (to the right). Even after peer interaction, 11% of groups continued to select option b, indicating that some TAs still struggled with applying the right-hand rule successfully.

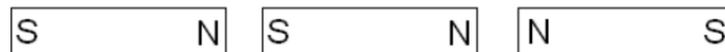

Fig B2.1: MCS item 1

Q1 tests if TAs can find out the direction of the net force on a central bar magnet placed between two identical bar magnets as shown in Fig. S3. As shown in Table I in the main text, the individual score for this question was 72%, which increased to 95% after group discussion, yielding an effect size of 0.57. Table II in the main text reveals that the most common distractor was option e ("zero"), followed by options a and c, each chosen by 5% of TAs. Despite this, the construction and co-construction rates for this question were 100%, indicating that most TAs successfully identified the correct answer through collaboration. Notably, one group still selected option e after peer interaction. This was the only instance where a group selected an incorrect answer when both students had given the correct answer individually. Explanations provided by upper-level students who selected option e suggest that some may have incorrectly assumed the net force was zero without fully analyzing the poles and the direction of the forces exerted by each magnet on the central one. One upper-level student explained, "Magnet at center is experiencing equal and opposite force".

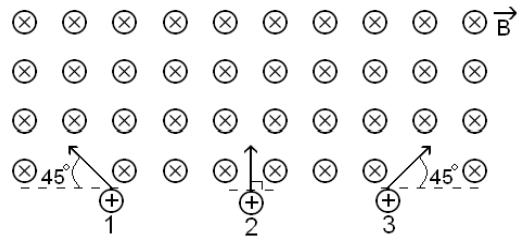

Fig B2.2: MCS item 14

In Q14, TAs analyzed a scenario where three identical positively charged particles enter a uniform magnetic field (into the page) at the same speed but in different directions, and they must determine which particle experiences the greatest magnetic force (Fig. S4). This question tests TAs' ability to visualize in three dimensions and calculate the magnetic force on a moving charged particle in the given magnetic field. Table I in the main text shows that Q14 had an effect size of 0.57 with individual and group scores of 72% and 95%, respectively. Table II in the main text indicates that 21% of TAs selected option b (2nd particle), while 5% chose option d (1st and 3rd particles). Some upper-level students who selected option b mistakenly claimed that the angle of 90 degrees between the second particle's velocity and the magnetic field maximizes the magnetic force, without realizing that all particles move in a plane perpendicular to the uniform magnetic field, and the angle between velocity and field is 90 degrees for all. Some upper-level students incorrectly factored in the specific angles made by the particles with the horizontal, thinking they influenced the magnetic force. For instance, one upper-level student incorrectly reasoned that "The angle between the velocity and the field is what matters here. Particle 2 has an angle of 90 degrees which will maximize the value of sin making the force the largest." However, peer interaction helped most groups identify the correct answer, showcasing the significant benefits of unguided group work.

Q7 challenged TAs to determine the direction of the magnetic field produced by a bar magnet at a point P, located inside the bar magnet equidistant from the two poles. This question assesses understanding of the internal magnetic field direction in bar magnets. Table I in the main text shows that Q7 had an individual score of 65%, which improved to 89% after group discussion, with a gain of 24% and an effect size of 0.56. According to Table II in the main text, 19% of TAs initially believed there is no magnetic field at the midpoint (option e), and 14% thought the field points toward the south pole (option c). Some groups selected options c or e even after collaboration. Upper-level undergraduates who selected option c commonly claimed that magnetic field lines always point from the north pole to the south pole, without considering the internal direction of the field. Others incorrectly likened magnetic poles to electric charges, leading to confusion. Despite this, most groups successfully reached the correct conclusion, demonstrating the value of peer discussions in clarifying alternative concepts.

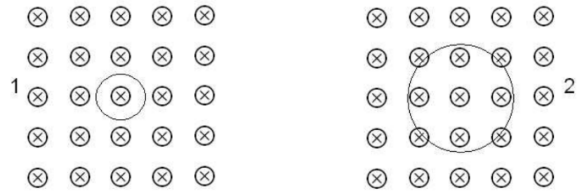

The figures 1 and 2 below show the circular paths of two different charged particles that travel at the <u>same</u> speed in <u>the same</u> uniform magnetic field $\vec{B}$ directed into the page. One particle is a proton and the other is an electron (which has a smaller mass than a proton).

The circle with the smaller radius in figure 1 is traveled by:

(a) the proton in a clockwise direction.
(b) the electron in a clockwise direction.
(c) the proton in a counterclockwise direction.
(d) the electron in a counterclockwise direction.
(e) Not enough information.

Fig B2.3: MCS item 20

Q20 assesses TAs' understanding of the motion of charged particles in a uniform magnetic field. The problem provides them with two diagrams showing circular paths of different radius, representing two charged particles traveling at the same speed within a magnetic field (directed into the page). Using Fig. S5 as a reference, they had to determine whether the smaller circular path corresponded to an electron or a proton and whether the motion was clockwise or counterclockwise. This question was the most challenging on the MCS survey, with the lowest individual and group scores. Table I in the main text indicates a 27% performance gain following unguided group work and an effect size of 0.56. The individual score was 47%, while the group score reached 74%, the lowest among all 30 questions. Table II in the main text shows that the correct answer, option b (electron in a clockwise direction), was selected by 47% of the TAs. However, 19% chose option d (electron in a counterclockwise direction), and 16% chose option e (not enough information). This item also had the lowest construction rate (70%). While most TAs correctly identified that the radius of the circular path is directly proportional to the mass of the particle, leading them to associate the smaller circular path with the electron, many struggled to determine the electron's direction of motion. These results suggest that while peer interaction improved understanding to some extent, significant challenges remained in visualizing and reasoning about the motion of charged particles in magnetic fields. For example, one upper-level student explained, "…we need to know the direction of the initial velocity to determine which direction the magnetic force acts in so we can determine the direction of motion for the electron. At any point, either cw or ccw directions make sense, but it just depends on the initial velocity to determine which one is actually the case."

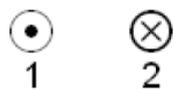

Fig B2.4: MCS item 22

Q22 investigates the direction of the magnetic force on a wire due to a parallel wire carrying current in the opposite direction (Fig. S6). Current in wire 1 flows out of the page and current in wire 2 flows into the page as shown in Fig. S6. TAs needed to determine the net force on wire 1 due to wire 2. Only 65% of TAs individually gave the correct answer (to the left) and 28% chose option d (to the right) as shown in Table II in the main text. After the peer interaction, 89% chose option c but 11% chose option d. From Table I in the main text, we can see that the effect size for this item is 0.56. Written explanations suggest that some upper-level students who chose the incorrect answer made an analogy with the opposite charges that attract each other. For example, one upper-level student choosing option d explained, "The wires create opposite B-fields, and thus should attract each other". Some used "opposites attract" incorrectly without considering the right-hand rule to determine the direction of force on wire 1 due to wire 2. Some of those who used the right-hand rule also ended up with an opposite direction of force which suggests that some of them did not visualize the directions properly and found it difficult to use the right-hand rule correctly. One upper-level student explained, "B travels clockwise around a wire with current going into the screen by the RHR [right hand rule]. Thus, the force would be attracting the wire 1 so it would be to the right."

Q29 asks about the direction of the magnetic field outside a circular loop in the plane of the loop carrying a clockwise current. It had an individual score of 65% which went to 89% after group work as shown in Table I in the main text. It had an effect size of 0.56 and a gain of 24%. From Table II in the main text, we see that the most common distractors were option a (into the page) and option e (the magnetic field is zero at point B). Both these options were chosen 14% of the time. The explanations from the upper-level students choosing option e show that some incorrectly thought that this set up is like that of a solenoid, so the magnetic field does not exist outside the loop. One upper-level student explained, "This is similar to a solenoid, it shouldn't have a field outside". Some upper-level students who chose option a considered that the field is directed into the page, for example, "same as above (Q28 which asks about the magnetic field inside the current loop) just that the net field strength is much less as much less of the current induces a field in the same direction".

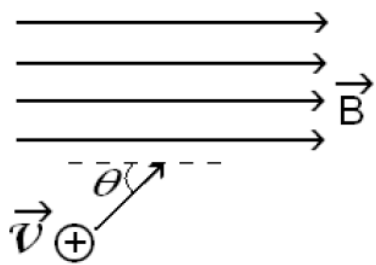

Fig B2.5: MCS item 16

Q16 states that a proton with speed v enters a uniform magnetic field of magnitude B making an angle $\theta$ with the direction of B as shown in Fig. S7. TAs had to describe the path of the proton after it enters the magnetic field. Table I in the main text shows that the group score was 89% with an effect size of 0.51. As shown in Table II in the main text, individually, about two thirds (67%) of TAs chose the correct answer A, which states that the proton has a helical motion with the pitch of the helix along the direction of the magnetic field. However, options b (Circular motion perpendicular to the plane of the page. The plane of the circle is perpendicular to the magnetic field, 9%) and option c (Circular motion perpendicular to the plane of the page. The plane of the circle is at an angle $\theta$ to the magnetic field, 23%) were also common choices. Options b and c were chosen by one group each after peer interaction. The written explanations suggest that some upper-level students only considered the component of velocity perpendicular to the magnetic field and did not take the parallel component into account while determining the path of the proton. For example, an upper-level student choosing b explained, "The particle experiences a force into the page when it enters the field and will eventually reach circular motion perpendicular to the B field." and another choosing c explained, "The magnetic field should produce circular motion perpendicular to the plane, but the plane of the circle would be slightly offset".

In Q27, TAs were presented with a diagram of two parallel wires both carrying current in the same direction. They need to determine the direction of the external uniform magnetic field if the net force on wire 1 is zero. Q27 investigates whether one can figure out the direction of the magnetic field if the direction of current in both wires and force on one of the wires are given. The most chosen option was a, into the page (the correct answer, 72%) followed by options e (none of the above, 14%) and b (out of the page, 5%) as shown in Table II in the main text. TAs reached a group score of 89% after peer interaction with an effect size of 0.42 as shown in Table I in the main text. But 11% of the groups chose option b showing that TAs had difficulty with three-dimensional visualization of the right-hand rule. The explanations of upper-level students who chose option b suggest that they may have had difficulty using the right-hand rule correctly. One upper-level student wrote, "The force on wire 1 from wire 2 is to the right. To counteract this, by the right-hand rule, a magnetic field will have to be coming out of the page." The written explanations also suggest that some students may have confused the direction of the magnetic field with the direction of the magnetic force. For example, one upper-level student explained, "The force on wire 1 from wire 2 is to the right and so in order to create a net magnetic force of zero the external field has to be left."

*B.3. Small Effect Size*

Table I in the main text highlights three items with small effect sizes ($d < 0.3$): Q9, Q12, and Q28. The smallest effect size ($d = 0.07$) was observed for Q9, which began with the highest individual score of 93% and increased slightly to a group score of 95%. Q9 assesses understanding of the forces acting on an electron entering a region with a uniform magnetic field at a velocity perpendicular to the field. The other two items, Q12 and Q28, each had an effect size of 0.27. Q12 focuses on the trajectory of an electron in a bubble chamber with a uniform magnetic field, starting with an individual score of 79% and improving to 89%. Q28 examines the direction of the magnetic field inside a current loop in the plane of the loop, with an individual score of 86% that increased to 95%. These small effect sizes reflect the fact that these questions were relatively easy for TAs, resulting in high individual scores that left little room for improvement through group interaction.